\documentclass[english,prl,aps,amsmath,amssymb,amsfonts,floatfix,superscriptaddr
ess,showpacs,twocolumn]{revtex4}
\usepackage[T1]{fontenc}
\usepackage[latin9]{inputenc}
\usepackage{verbatim}
\usepackage{float}
\usepackage{graphicx}

\makeatletter

\usepackage{epsfig}

\def\beq{\begin{equation}}
\def\beqno{\begin{equation}\nonumber}
\def\eeq{\end{equation}}

\def\li6{$^6$\rm{Li}}

\def\beq{\begin{equation}}
\def\eeq{\end{equation}}

\makeatother

\usepackage{babel}

\begin{document}

\title{Local moment versus Kondo behavior of the $4f$-electrons\\
in rare-earth iron oxypnictides}

\author{Leonid Pourovskii}
\affiliation{Centre de Physique Th\'{e}orique, \'{E}cole Polytechnique, CNRS, 91
128 Palaiseau,
France}
\author{Ver\'{o}nica Vildosola}
\affiliation{Centre de Physique Th\'{e}orique, \'{E}cole Polytechnique, CNRS, 91
128 Palaiseau,
France}
\affiliation{Departamento de F\'{\i}sica, Comisi\'{o}n Nacional de Energ\'{\i}a
At\'{o}mica (CNEA-CONICET),
Provincia de Buenos Aires, San Mart\'{\i}n, Argentina}
\affiliation{Japan Science and Technology Agency, CREST}
\author{Silke Biermann}
\affiliation{Centre de Physique Th\'{e}orique, \'{E}cole Polytechnique, CNRS, 91
128 Palaiseau,
France}
\affiliation{Japan Science and Technology Agency, CREST}
\author{Antoine Georges}
\affiliation{Centre de Physique Th\'{e}orique, \'{E}cole Polytechnique, CNRS, 91
128 Palaiseau,
France}
\affiliation{Japan Science and Technology Agency, CREST}

\begin{abstract}
We consider the role played by the $4f$ states in the rare-earth oxyarsenides REOFeAs (RE=Ce,Pr,Nd) and
the oxyphosphate CeOFeP, using a first-principles technique that combines the local density approximation
and dynamical mean-field theory (LDA+DMFT).
In the Pr and Nd compounds, the 4$f$ states are located well below and above the Fermi
level $E_F$, and essentially do not interact with the iron 3$d$ bands located near $E_F$,
resulting in local moment behavior.
In the Ce compounds, our results reveal a qualitatively different picture,
with the 3$d$-4$f$ hybridization being sufficiently strong to give rise to an
observable Kondo screening of the local 4$f$ moment.
Our LDA+DMFT electronic structure calculations allow us to estimate the Kondo temperature $T_K$
for both CeOFeP and CeOFeAs. For the phosphate, the order of magnitude of our estimate is
consistent with the experimental observation of $T_K\simeq 10$K.
At ambient pressure, $T_K$
is found to be negligibly small for CeOFeAs. Under applied hydrostatic pressure, we predict an
exponential increase of $T_K$ which reaches values comparable to the superconducting
$T_c \simeq$40~K at pressures above $10$GPa. We conjecture that the competition
between the Kondo effect and superconductivity may be at the origin of the monotonous decrease
of $T_c$ observed in CeOFeAs under pressure. We argue that the quantitative aspects of this
competition are inconsistent with a weak-coupling BCS
description of the superconductivity in the oxyarsenides.
\end{abstract}

\maketitle

The discovery of the
REO$_{1-x}$F$_x$FeAs superconductors with a
transition temperature up to $T_c$= 55 K
\cite{Kamihara-LOFA,Zhi-AnRen,gchen,ren,ren2,yang}
and the observation of superconductivity
also in oxygen deficient REFeAsO$_{1-\delta}$\cite{ren3}
has triggered enormous efforts, both experimentally and
theoretically, aimed at understanding the electronic properties
of rare earth iron oxypnictides.
The evolution of $T_c$ along the series of rare earth
elements, and possible correlations with structural
properties have attracted particular interest.
While LaO$_{1-x}$F$_x$FeAs has a $T_c$ of 26 K,
the critical temperature is drastically increased
by replacing La by other rare earth ions
(RE=Ce,Pr,Sm,Nd,Gd),
up to $\sim$55K for Sm .
Possible correlations between the evolution
of $T_c$  with  changes in the structural
parameters due to decreasing size of the rare-earth ions along the series have been reported
in the literature \cite{ren3,zhao}.

On the theoretical side,
most {\it ab initio} studies have so far concentrated on the La compound.
To our knowledge, only the work of
Nekrasov {\it et al.} \cite{nekrasov} has addressed the question of how
 the electronic structure of REOFeAs changes upon rare-earth substitution.
On the basis of LDA calculations,
these authors obtained essentially identical electronic structures
for all the REOFeAs compounds considered (with RE=La,Ce,Pr,Nd,Sm,Y).
In their work [\onlinecite{nekrasov}], the 4$f$ shells of the rare-earths were treated
 as core states. However, an intriguing question is whether the low-energy electronic structure
 of REOFeAs might be modified along the rare earth series, due to the hybridization between the Fe 3$d$ band and the
 localized 4$f$ states of the rare earth ions.
 This possibility has not been so far investigated in the literature. Meanwhile, a Kondo screening
 of the Ce local magnetic moment on the 4$f$ shell has been observed in the CeFePO compound,
 an homologue of CeFeAsO, with a reported Kondo temperature $T_K=10$~K \cite{bruning}.
 This implies the existence of an hybridization between the Fe 3$d$ bands in the vicinity of
  $E_F$  and the localized Ce 4$f$  states.
The discovery of the Kondo effect in CeFePO raises the possibility that the
Re$4f$-Fe$3d$ hybridization might similarly lead to a Kondo screening of the
RE local moment in some of the oxyarsenide compounds.

In order to investigate the 4$f$ states and their interaction with other bands, a many-body treatment
of the strong Coulomb interaction between 4$f$ electrons is necessary.
Therefore we have employed the combined local density approximation
and dynamical mean-field theory (LDA+DMFT) approach \cite{dmft_rev,lda_dmft}
to study properties of 4$f$ states in CeOFeP and REOFeAs (RE=Ce,Pr,Nd).
We show that in the Pr and Nd compounds the 4f states are very
localized, leading to unscreen local moments.
The occupied 4f states are located well
below the 3d-Fe states,
and the unoccupied ones well above.

We find that the situation is different, however, for the
Ce compounds CeOFePn (Pn=P and As), for
which the occupied $4f$ band is located at the bottom
or just below the bottom of the 3d-Fe states. The resulting hybridization
leads to Kondo screening of the 4$f$ Ce local moment at low T.
We estimate the Kondo temperatures of
CeOFeP, and find that the order of magnitude is consistent with experiments.
We predict a rapid increase of the
(very low) Kondo scale of CeOFeAs as a function of pressure,
due to the contraction of the Fe-Ce interatomic distance and
to the corresponding increase of the $f$-$d$ hybridization.
Moreover, we conjecture that the Kondo effect in CeOFeAs may be at the origin
of the rather rapid suppression of $T_c$ under pressure
observed in doped CeOFeAs \cite{zocco} (which is in sharp contrast
to the behaviour under pressure of LaOFeAs~\cite{zocco} at a
similar doping level).

We have employed the LDA+DMFT approach using the recently developed
fully-selfconsistent
framework described in \cite{pour2007}.
%
%
The local self-energy of the $4f$ shell is computed according to the DMFT
prescription and by employing the atomic (Hubbard-I) approximation
\cite{hub1}.
This approach to local correlations
has been shown to be appropriate for the localized 4$f$ shells of
rare-earths compounds \cite{pour2007}.
 We have used the full four-index $U$ matrix for the
 local Coulomb interaction. The spin-orbit coupling and splitting
of the bare 4$f$ levels due to the crystal field
 were taken into account. Our calculations
are performed for the paramagnetic state and at zero temperature.

 In order to determine the value of the local Coulomb interaction $U$ on the 4$f$ shell,
 we have performed constrained LDA calculations.
 We obtain a value of about $9.7$~eV for both CeOFeAs and CeOFeP.
 This is substantially larger than the usual range of $U$ values for pure Ce.
 This result can be explained by the quasi two-dimensional environment of the rare-earth sites
 in the case of oxypnictides, with only four nearest-
 neighbours present, which leads to a rather poor screening of the local
 Coulomb interaction.
 Values of the Slater integrals $F^2$, $F^4$,
 and  $F^6$, which are known to be weakly dependent on crystalline environment,
 have been taken from the optical measurements of \cite{carnal}.
The corresponding values of the exchange parameter $J$ are equal
 to be $0.69$, $0.73$, and $0.77$~eV for Ce, Pr, and Nd, respectively.
 For the double counting correction, we have employed the fully-localized limit expression
 $U(N_f-1/2)-J(N_f/2-1/2)$, where the occupancy $N_f$ in the atomic limit
 and at zero temperature is equal to 1, 2, and 3 for Ce, Pr, and Nd,
 respectively.

All calculations have been carried out at the experimental lattice parameters
$a$ and $c$ \cite{gchen,ren,Zhi-AnRen}. An unusual sensitivity of the low-energy
electronic structure of iron oxypnictides to the vertical distance between the Fe and As planes $z_{As}$
 has been pointed out in the literature \cite{mazin2,we}. However, in the present work,
 we mainly focus on high-energy features of the electronic
structure, particularly on the 4$f$ bands and their interaction with other bands.
Those features are not expected to be very sensitive to small changes in
the Fe-As distance, they may rather show some sensitivity with respect to the vertical
position of the rare-earth plane $z_{RE}$, which is almost constant
along the series \cite{zhao,POFA,chulee}. Thus the experimental values of
the internal parameters $z_{As}$ and $z_{RE}$ for LaOFeAs \cite{cruz} have been used
for all oxyarsenides studied, while CeOFeP has been computed with the internal parameters fixed at their
experimental values for PrOFeP \cite{zimmer}.

Because the atomic Hubbard-I approximation to strong correlations
is not able to capture directly the Kondo effect, we have employed the approach
of Gunnarsson and Sch\"onhammer \cite{gunn} to calculate the Kondo
temperature. Within this approach the Kondo temperature is given
(in the Kondo regime) by the following expression:
\beq
\label{TK_eq}
T_K=D e^{-\frac{|\epsilon_f|}{N_f \rho_0 V_{cf}^2}},
\eeq
where $\epsilon_f$ is the average position of the occupied 4$f$ level, $\rho_0$ is
the density of states at the Fermi level, $N_f$ is the $f$ band degeneracy, $V_{cf}$
is the average hybridization matrix element between the 4$f$ level and the conduction band,
$D$ is the bandwidth of the occupied part of the conduction band.
As shown in \cite{gunn},
the spin-orbit splitting yields a rather small correction to $T_K$ and can be neglected, so that
we take $N_f$=14. The value of $V_{cf}^2$ can be estimated from the imaginary part
of the DMFT hybridization function at $E_F$, which reads:
\beq
\mathrm{Im} \Delta(E_F)=
\mathrm{Im} \sum_{\bf k}\frac{|V^{\bf k}_{cf}|^2}{E_F-\epsilon_{\bf k}} \approx \pi \rho_0 V_{cf}^2,
\eeq
where $\epsilon_{\bf k}$ are the conduction band states.
For a degenerate case, the average value of $V_{cf}^2$ can  be extracted from
$\mathrm{Tr\,Im} \Delta(E_F)/N_f$. An accurate evaluation of the parameters entering in the
exponential factor in (\ref{TK_eq}) is necessary to obtain
any reasonable estimate of $T_K$. Thus we have employed a 9000 $\bf{k}$-point mesh in the Brillouin zone in our LDA+DMFT calculations in order
to evaluated the average hybridization $V_{cf}$ and the 4$f$ level position $\epsilon_f$. 
To obtain an accurate value of the density of states at the Fermi level, we use
a Full-Potential APW+local orbitals code\cite{wien2k} with the 4$f$ states in the core.
We have checked that all the values used in calculation of $T_K$ are converged with respect to the ${\bf k}$-point mesh.

\begin{figure}
\includegraphics[width=0.85\columnwidth]{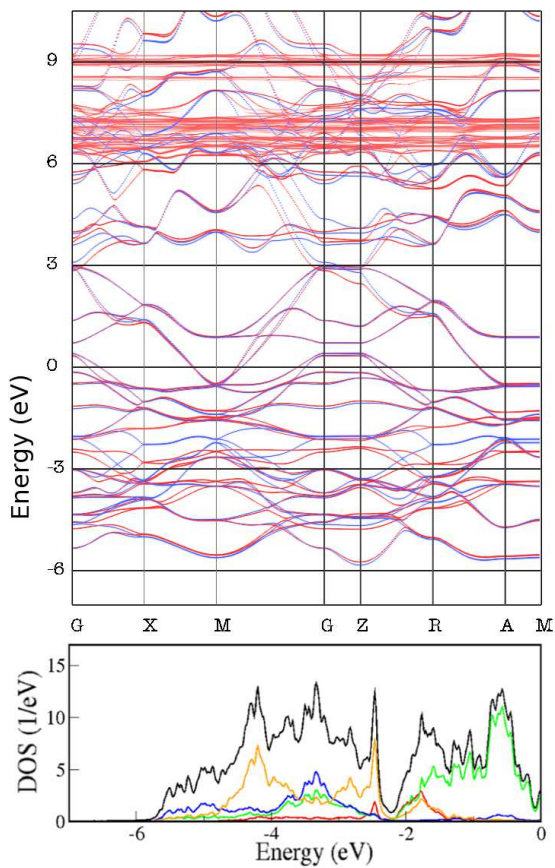}
\caption{\label{CeOFeP_bands}
Electronic structure of CeOFeP obtained within LDA with the 4$f$ states treated as core (blue line) and within LDA+DMFT (red line). The 4$f$ states of RE form a rather dispersionless upper Hubbard band 6-9 eV above the Fermi energy. The lower Hubbard band is more dispersed due to hybridization with O 2$p$ and Fe 3$d$. It is located between -1.8 and -2 eV/ In the lower panel the density of occupied states is displayed, with the total, partial Ce 4$f$, Fe 3$d$, pnictogen $p$ and O 2$p$ shown
by black, red, green, blue, and orange curves, respectively. One may notice a rather strong hybridization between the O 2$p$ states at the top of $p$ band and the occupied 4$f$ states.}
\end{figure}

First, we discuss the LDA+DMFT electronic structure of the CeOFeP compound in connection
with the Kondo screening of the Ce local magnetic moments observed experimentally in this compound.
In the upper panel of Fig.~\ref{CeOFeP_bands} the LDA+DMFT spectral function
 at $U-J=$9 eV (in red) is  superimposed to the band structure obtained by treating the
 Ce 4$f$ states as core (in blue). The empty Ce 4$f$ states (the upper Hubbard band) form rather
 dispersionless bands between  6 and 9 eV above $E_F$  hybridized to a certain degree with the
Ce 5$d$ band. The occupied part (the lower Hubbard band) of the Ce 4$f$ states
is located at about 1.8-2 eV below $E_F$. It hybridizes substantially with the $p$ states
of mostly oxygen character at the top of the As/O $p$ band, causing a downward shift of
0.5 eV of the oxygen 2$p$ band that was located at $\sim$ 2.3 eV in the LDA with f in core
 band structure. In the lower panel the corresponding partial density of (occupied) states
 is depicted.
Note that, because the Hubbard-I approximation does not capture the Kondo peak,
the LDA+DMFT (Hubbard-I) spectral function in Fig.~\ref{CeOFeP_bands} should be
viewed as the band structure for $T>T_K$.

From the  imaginary part of the Ce 4$f$ hybridization function we have obtained $V_{cf}$=116 meV,
while $\epsilon_f$=-2.0 eV, $\rho_0$=1.85 (eV$\times$formula unit)$^{-1}$, and $D \sim $2~eV for the Fe 3$d$ bands,
resulting in an estimate of $T_K\simeq 77$K. This is a reasonable order of magnitude in
comparison to the experimental result $T_K$=10 K of Br\"uning {\it et al.} \cite{bruning}, given
that our estimate for $T_K$ is obtained for the single
impurity case. Indeed, the corresponding $T_K$ for the Kondo lattice is expected to be reduced by
at least a factor of $2$ for the approximately half-filled conduction (Fe 3$d$) band under consideration\cite{Burdin}.
Our estimates for $\epsilon_f$ and $\rho_0 V_{cf}^2$=25 meV also compare well with the corresponding values of -2.4 eV and 19 meV, respectively,
 obtained for CeCu$_2$Si$_2$ by Kang {\it et al.} \cite{kang} from a fit to the experimental PES spectra.
 As noted in Ref.~\cite{bruning}, CeCu$_2$Si$_2$ and CeOFeP have about the same Kondo
 temperature and  Sommerfeld coefficient $\gamma$ of the specific heat.
\begin{figure*}
\includegraphics[width=1.9\columnwidth]{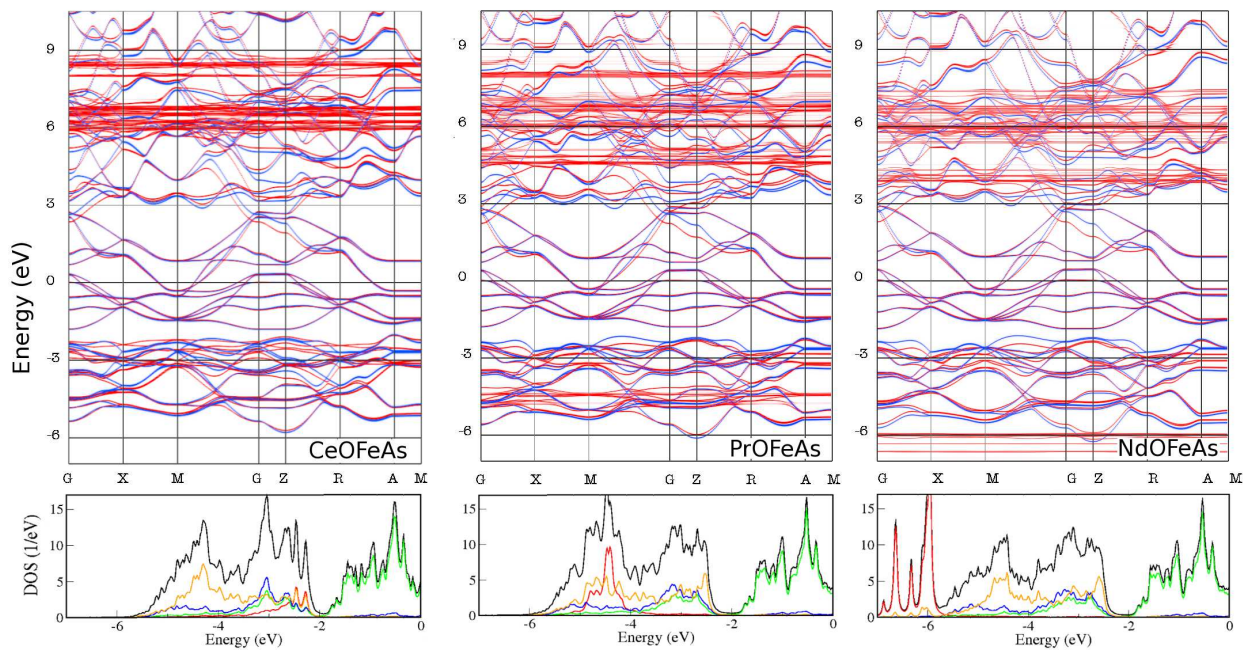}
\caption{\label{ReOFeAs_bands}
Electronic structure of stoichiometric iron oxyarsenides obtained within LDA with the 4$f$ states treated as core (blue line) and within
LDA+DMFT (red line). The empty 4$f$ states of RE form a rather dispersionless  upper Hubbard band significantly above the Fermi energy.
The lower Hubbard band is located just below the bottom of the Fe 3$d$ band, in the middle of the O/As $p$ band and below the $p$ band
in CeOFeAs, PrOFeAs, and NdOFeAs, respectively. The lower panel displays the corrseponding densities of states, the color coding is the same
as in Fig.~\ref{CeOFeP_bands}}
\end{figure*}

We have also performed LDA+DMFT calculations for CeOFeP using the "standard" value of 6~eV for
the $U$ parameter of pure Ce 4$f$. Those calculations predict
the lower Hubbard band to be much closer to the Fermi level, with $\epsilon_f\simeq -0.5$eV.
This would lead to a unrealistically high Kondo temperature for CeOFeP, even if one takes into
account necessary corrections to (\ref{TK_eq}) due to deviation from the Kondo regime
(mixed valence behavior) for small values of
$|\epsilon_f|$. Therefore, the constrained LDA value of the local Coulomb interaction
$U-J$=9 eV  provides a better description of the Kondo behaviour in CeOFeP
as compared to the usual choice of $U$ for pure Ce. We used the same
value to study the oxyarsenides.

In Fig.~\ref{ReOFeAs_bands}, we compare the LDA+DMFT band structures of
stoichometric CeOFeAs,  PrOFeAs and NdOFeAs with the band structures obtained for
the same compounds within LDA by treating the rare-earth 4$f$ states as core. Differences between the LDA+DMFT and LDA-with-4$f$-in-core electronic structures are
especially pronounced well above $E_F$, where the unoccupied 4$f$ states form a set of dispersionless bands in the range of energies from 6 to 9 eV, from 4 to 8 eV,
and from 3 to 7 eV for the Ce, Pr, and Nd compounds, respectively. The width of about 3 eV of those upper Hubbard bands is due to multiplet splittings associated with the exchange and spin-orbit interactions.
The occupied 4$f$ states are located in the range of energies from -2.3 to -3 eV in CeOFeAs, from -4 to -5 eV in PrOFeAs, and below -6 eV for NdOFeAs. Similarly to the
case of CeOFeP, the occupied 4$f$ states hybridize rather strongly with the oxygen $p$ states, which are located mainly at the top of the As 4$p$/O 2$p$ band. This
4$f$-2$p$ hybridization leads to some modifications at the top of the O/As $p$ band in comparison with the $f$-in-core
band structure. In contrast, the Fe 3$d$ bands
are hardly modified by the interaction with the 4$f$ states, as one may see in the
upper panel of Fig.~\ref{ReOFeAs_bands} where blue and red bands corresponding to Fe (close to the Fermi level)
essentially coincide.
From these results, we conclude that the $4f$ electrons behave as unscreened localized moments in
the Pr- and Nd- compounds.

In contrast, for CeOFeAs, the lower Hubbard band is located
rather close to the Fe 3$d$ bands (see the corresponding DOS on Fig.~\ref{ReOFeAs_bands}). Thus,
in analogy to the case of CeOFeP, in CeOFeAs one might expect to observe a Kondo screening
of the 4$f$ local moment. We have estimated
the Kondo temperature in CeOFeAs using formula
(\ref{TK_eq}), and the resulting values are given in the first row of
Table~\ref{table:par_vs_P}.
With $D\simeq 2$eV for the Fe$3d$ bands, this leads to a value of $T_K$ of order
10$^{-4}$~K. Thus, $T_K$ in CeOFeAs is negligible in comparison to the magnetic ordering
temperature for the Ce local moments, which was measured to be about 5$~K$ in
this compound\cite{gchen}. This drastic reduction of the Kondo scale (by about 5 orders of magnitude !)
in comparison to $T_K$ for CeOFeP is mainly due to a substantially weaker $3d$-$4f$ hybridization
in CeOFeAs. This decrease of the $d$-$f$ hybridization is rather expected, as the Fe-Ce distance
is larger in CeOFeAs than in CeOFeP.

In order to investigate the behavior of the Kondo scale under applied pressure we have estimated the corresponding
change in volume of CeOFeAs. Methods based on the atomic sphere approximation can not reliably describe
elastic properties of a complex crystal structure.
One may notice that the structural parameters of CeOFeAs and LaOFeAs are rather similar, while the 4$f$ states of Ce,
being still essentially localized,
are not expected to affect substantially the elastic properties of CeOFeAs.
Therefore, for CeOFeAs we have adopted the theoretical elastic constants computed for LaOFeAs \cite{shein} within a full-potential approach.
With those elastic constants the relative values of the lattice parameters $a/a_0$($c/c_0$)
of CeOFeAs (where $a_0$ and $c_0$ are the corresponding
values at zero pressure) are equal to 0.985(0.978), 0.970(0.956), and 0.955(0.934)
under hydrostatic pressure of 5, 10, and 15 GPa, respectively.
Anisotropy of the elastic constants leads to a substantially larger relative contraction
for the $c$- parameter compared with the one for $a$.

\begin{table}
\caption{\label{table:par_vs_P}
The average hybridization $V_{cf}$ and 4$f$ level position $\epsilon_f$, density of states at the
Fermi level $\rho_0$, and Kondo temperature $T_K$ in CeOFeAs as function of applied pressure
}
\begin{ruledtabular}
\begin{tabular}{lcccc}
P(GPa) & $V_{cf}$ (meV) & $\epsilon_f$ (eV)  & $\rho_0$ (eV*formula unit)$^{-1}$ & $T_K$ (K) \\
\hline
0 & 77 & -3.03 & 2.0 & 0.0003 \\
5 & 90 & -2.74 & 1.86 & 0.067 \\
10 & 111 & -2.38 & 1.77 & 9.8 \\
15 & 129 & -2.04 & 1.70 & 137.0 \\
\end{tabular}
\end{ruledtabular}
\end{table}

In Table~\ref{table:par_vs_P} we list, as a function of applied pressure,
 the values of the average hybridization $V_{cf}$ and 4$f$-level position $\epsilon_f$
obtained within LDA+DMFT as well as the density of states at the Fermi level.
One may see that $V_{cf}$ grows substantially with increasing pressure, while the 4$f$-level position is
shifted towards the Fermi level. The reduction of the
Ce-Fe distance naturally leads to an increase in the corresponding hybridization matrix elements,
the contraction in the Ce-O distance
further enhances the strong O 2$p$-Ce 4$f$ hybridization, leading to the upwards shift of $\epsilon_f$. The Fe 3$d$ bandwidth
increases as well under pressure, leading to a drop in $\rho_0$.
However the effects due to the sharp rise in  $V_{cf}$ are obviously the most important, leading to a reduction of
$|\epsilon_f|/(\rho_0 V_{cf}^2)$ in formula (\ref{TK_eq}) and to a corresponding (almost exponential) increase of the Kondo temperature
as a function of pressure (see Table~\ref{table:par_vs_P}).
Hence, $T_K$ raises from the negligible value of $10^{-4}$~K at ambient pressure
to a value of order $100$~K under an applied pressure of $15$~GPa.

\begin{figure}
\includegraphics[width=0.9\columnwidth]{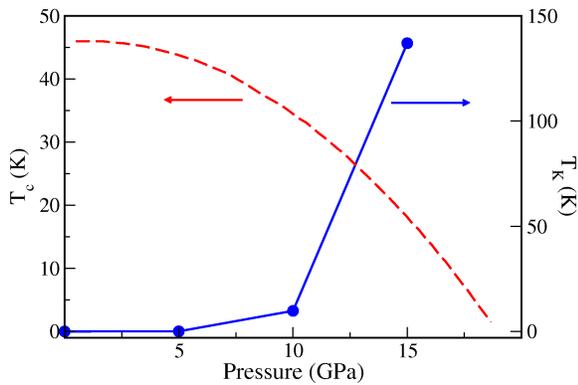}
\caption{
\label{Tc_vs_TK_P}
Experimental evolution of the superconducting transition temperature $T_c$ in CeO$_{0.88}$F$_{0.12}$FeAs
under applied pressure\cite{zocco}(red dashed curve) and our theoretical estimate for the corresponding
evolution of the Kondo temperature $T_K$ in CeOFeAs (blue solid curve).}
\end{figure}

Within our approach one may only obtain a rather crude estimate for $T_K$.
But our {\it qualitative conclusions} should be rather robust: as the pressure increases
the Kondo temperature exhibits an exponential growth, from
values which are negligible compared to the superconducting transition temperature $T_c$
at ambient pressure, up to values of the same order of magnitude as $T_c$ at pressures
above $10$GPa. It is interesting to contrast
the predicted enhancement of $T_K$ under pressure with the behavior of $T_c$ observed
experimentally for CeO$_{0.88}$F$_{0.12}$FeAs \cite{zocco} (see Fig.~\ref{Tc_vs_TK_P}).
In contrast to LaO$_{x}$F$_{1-x}$FeAs at a very similar doping level, for which $T_c$ {\it raises}
with increasing pressure up to 5 GPa before starting to decrease rather slowly \cite{zocco},
in CeO$_{0.88}$F$_{0.12}$FeAs one observes a monotonous
and rather steep decrease of $T_c$ with increasing pressure.
The superconducting transition is actually completely suppressed at pressures of about
20 GPa. As one may see in Fig.~\ref{Tc_vs_TK_P}, a rapid reduction of $T_c$ sets in at
pressures (above 5 GPa), at which $T_K$ starts
to reach values of order of a few degrees Kelvin. Thus, we may conjecture that the
rapid suppression of superconductivity observed in the doped
CeOFeAs under pressure is due to the corresponding stabilization of
a competing heavy-fermion phase with Kondo-screened local moments of the Ce shell.

The competition between the Kondo effect and $s$-wave superconductivity has been
investigated theoretically in Ref.~\cite{barzykin}, in the framework of the periodic
Anderson model.
This study concluded that with $T_K > T_{c0}^2/W$ (where $T_{c0}$ is the transition
temperature in the absence of the Kondo effect and $W$ is the bandwidth of the conduction band),
the $s$-wave superconductivity becomes rapidly suppressed, with $T_c$ decreasing exponentially
as a function of $T_K$.
This expresses the competition between the Kondo screening energy ($\sim T_K$) and the
condensation energy $\sim T_{c0}^2/W$ of a weak-coupling BCS superconductor.
In our case the threshold value of $T_K=T_{c0}^2/W$ is of order of 0.1~K. However, our
estimates of $T_K$ lead us to conclude
that while the Kondo effect may be responsible for a rather rapid suppression of
the superconductivity in CeO$_x$F$_{1-x}x$FeAs under applied pressure,
the actual superconducting phase is still much more robust than what would be predicted by the
theory in Ref.~[\onlinecite{barzykin}]. This points out to important differences
between the nature of the superconductivity of the rare-earth iron oxyarsenides
and that of a weak-coupling BCS superconductor.

In conclusion, our study reveals that while for the heavier rare-earths the electrons in
the $4f$ shell of rare-earth oxyarsenides behave as unscreened local moments, the
Cerium-based compounds behave in a different manner. There, a competition between
Kondo screening and superconductivity takes place under applied pressure. This may be
responsible for the suppression of superconductivity under applied pressure in the
cerium-based compounds, and suggests that the competing phase at large pressure should be
a heavy-fermion state with a moderate effective-mass enhancement.
Orders of magnitude suggest however that the Kondo effect is less detrimental to
superconductivity in these compounds than for a weak-coupling BCS superconductor, hence
providing indirect evidence for the unconventional nature of superconductivity
in the oxyarsenides. This competition between superconductivity and a heavy-fermion state
deserves further theoretical and experimental investigations.

\acknowledgements
We are grateful to J.Bobroff, Paul C.W. Chu, L.Pinsard-Gaudart and
A.Revcolevschi for useful discussions. This work was supported by IDRIS Orsay
under project number 081393.

\end{document}